\documentclass{vgtc}                          

\ifpdf
  \pdfoutput=1\relax                   
  \usepackage{graphicx}                
  \DeclareGraphicsExtensions{.pdf,.png,.jpg,.jpeg} 
\else
  \usepackage{graphicx}                
  \DeclareGraphicsExtensions{.eps}     
\fi%

\graphicspath{{figures/}{pictures/}{images/}{./}} 

\usepackage{microtype}                 
\PassOptionsToPackage{warn}{textcomp}  
\usepackage{textcomp}                  
\usepackage{mathptmx}                  
\usepackage{times}                     
\usepackage{cite}                      
\usepackage{tabu}                      
\usepackage{booktabs}                  
\usepackage{enumitem}

\onlineid{0}

\vgtccategory{Research}

\vgtcinsertpkg

\title{Pulse: Toward a Smart Campus by Communicating Real-time Wi-Fi Access Data}

\author{Aoyu Wu\thanks{e-mail: \{awuac, bkku, fchengaa, xshuaa, apuri, ywangjh, huamin\}@ust.hk} %
\and Bon Kyung Ku\footnotemark[1]
\and Furui Cheng\footnotemark[1]
\and Xinhuan Shu\footnotemark[1]
\and Abishek Puri\footnotemark[1]
\and Yifang Wang\footnotemark[1]
\and Huamin Qu\footnotemark[1]}
\affiliation{\scriptsize The Hong Kong University of Science and Technology}

\teaser{
  \includegraphics[width=1.0\textwidth]{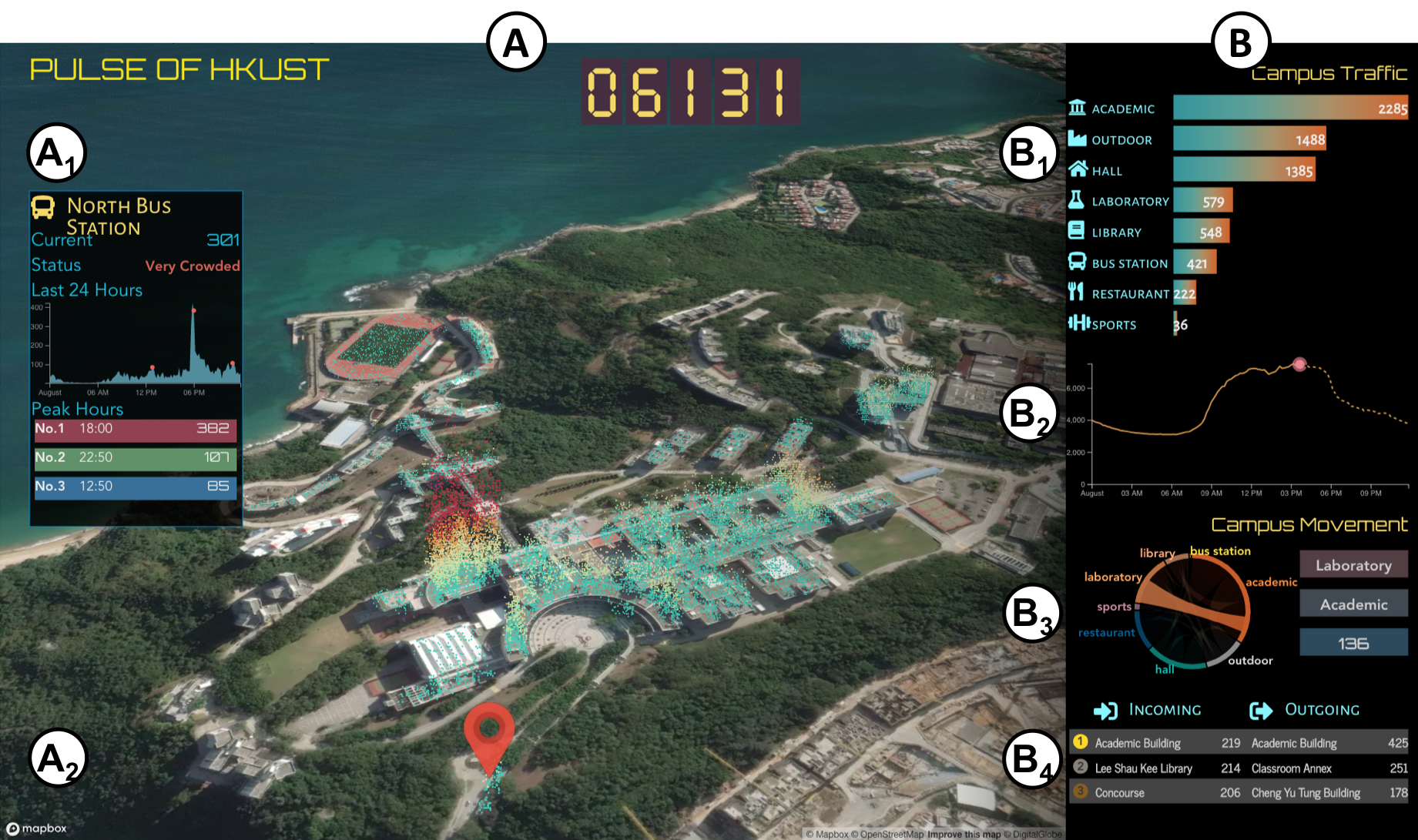}
  \caption{The Pulse interface communicates the campus crowd information to the lay public through a projection screen: (A) The Map View provides a holistic view of the crowd distribution on the satellite map ($A_2$), with a pop-up window ($A_1$) displaying the derived information of campus facilities in a loop; (B) The Chart View presents the statistical information including the crowd amount ranking of functional zones ($B_1$), the historical and predicted number ($B_2$), the movement of crowd among functional zones ($B_3$), and the buildings with top incoming and outgoing crowds ($B_4$).}
}

\abstract{To enhance the mobility and convenience of the campus community, we designed and implemented the Pulse system, a visual interface for communicating the crowd information to the lay public including campus members and visitors. This is a challenging task which requires analyzing and reconciling the demands and interests for data as well as visual design among diverse target audiences. Through an iterative design progress, we study and address the diverse preferences of the lay audiences, whereby design rationales are distilled. The final prototype combines a set of techniques such as chart junk and redundancy encoding. Initial feedback from a wide audience confirms the benefits and attractiveness of the system.} 

\CCScatlist{
  \CCScat{H.5.m}{Information Interfaces and Presentation}{}{}{}
}

\begin{document}


\firstsection{Introduction}

\maketitle

With the ever-growing scale of higher education, crowdedness in campus facilities arises as a major issue which causes inconvenience and hinders mobility. However, this could be palliated if people are provided with crowd information for making decisions on mobility. Therefore, we have tried various approaches for sensing and visualizing the crowd data on campus. Among various data resources, the Wi-Fi access log information stands out because of its availability and reliability \cite{Binthaisong:2017:WiCrowd}.

With this capacity comes the task for designing visualization systems for communicating the crowd information to the audiences. Specifically, we were asked to develop such communicative visualization on the projection screens within campus. This is challenging due to the difficulty in conveying the huge amount of data on a non-interactive interface. Furthermore, it is non-trivial to address the diverse audiences who vary in demands and interests of both data and visualizations.

Through an iterative design process - via individual interviews, paper- and code-based prototyping - we studied and distilled the design rationales for communicating crowd information to the lay audiences. Specifically, we observed the difference among audiences' preferences for visual formats including texts and charts. We also found that audiences tended to attract attention by competitive data such as rankings. In addition, we took a look at how redundant encoding and animation could promote effective communication. We believe that our lessons could contribute to the continued research in similar application areas for visualizing crowd information.

\section{Related Work}
Much work has been proposed to analyze the Wi-Fi access data for behavior analysis \cite{Zeng:2015, Prasertsung:2017} or facility management \cite{Kergaard:2012:Challenge}. However, there were few visualization systems proposed to communicate those information. A typical visualization approach is the heat map \cite{Horwitz:2018, Isaac:2017}. Other systems display crowd sizes at different locations using circles on a 2D map \cite{Ahlers:2016}. These approaches could  deliver spatial distribution information, but could lack in sense of dynamic and realistic because the 2D maps do not provide intuitive representation of different locations. In addition, the exclusion of statistical information could make it impractical for real-life usage.

To address those issues, a recent system called Wi-Crowd \cite{Binthaisong:2017:WiCrowd} visualizes real-time crowd distribution on campus using a 3D building map, and a 2D information panel which includes useful numerical information such as top buildings of this hour. While it is shown to be useful and engaging, it remains unclear how those decisions on data and designs are made and whether the approaches are effective. For instance, 3D models and design used in the project were not sufficiently representative of the different locations on campus and therefore, showing weakness in intuitive information delivery.

Our work adds to this conversion by contributing a set of design rationales that guide communicative visualization for this application area. Meanwhile, we reported on the decision choices that we made based on successful or failed prototypes, providing anecdotal evidence on several visualization techniques that might be helpful for such system.

\section{Design Rationales}
Our first contribution is the derivation of design rationales for systems which communicates Wi-Fi access data. The goal of the Pulse system is to analyze the Wi-Fi access data and communicate the derived useful information to the university community and visitors, promoting the idea of smart campus. 

Through the design progress, we have worked with a variety of audiences including 4 administration staffs, 10 undergraduate students, 14 postgraduate students, and 6 external personnel. We adopted different design approaches including individual interviews where we directly asked for their interests, paper- and code-based prototyping where we gathered feedback. In the following text, we summarize the design rationales that we distilled during the iterative design progress:

\textbf{R1: Data visualization must come with geographical context.} Audiences need to be cognizant of the geographical locations to enhance the sense of realistic and engagement. This is particularly important for visitors which are unfamiliar with the environment.

\textbf{R2: Maintain visual redundancy by dual-modality of text and visualization.} We found that audiences tended to fall into two categories: those who preferred graphical charts and those who liked textual information such as tables. We believe that both types are valid and our system shall address them by redundancy.

\textbf{R3: Adopt simple and intuitive charts.} We observed an overall negative altitude towards complex visualization. Therefore, the system shall adopt simple, daily-used charts. Complex visualization shall emerge step-by-step to promote understanding.

\textbf{R4: Involve competitive information.} We noted that audiences tended to get attracted by competitive information such as rankings. We think the representation of such information could enhance the attractiveness of the communicative system.

\section{Visualization System}
Our second contribution is the development of the Pulse system, which visualizes the Wi-Fi access data with a 3D Map view and derived useful statistical information with a 2D Chart view. In the following text, we described each visual component and explained the design decisions.

\subsection{Map View}
The Map View (Figure 1.A) consists of two main visual components: a satellite map which offers an overview of the crowd distribution calculated from the Wi-Fi access data, and a pop-up window which provides useful information of important facilities in a loop.

\subsubsection{Satellite Map}
We utilize the point cloud on the satellite map (Figure 1.$A_2$) to provide the overall information (\textbf{R1}). Specifically, the height of the point cloud indicates the number of Wi-Fi connections at each building. We have made two improvements based on the feedback to improve the readability: First, we apply bouncing effect to the point cloud, so that the points are vibrating from the 80\% to 120\% of original heights. This could bring about a sense of dynamic and vitality, which our audiences appreciated much. However, such changing heights could lead to misconception of the actual value. Regarding this, our second improvement is to assign a linear gradient colorway. For instance, the cyan color denotes less than 200 connections, while the red color corresponds to more than 1000 connections. The colorway is designed based on the \textbf{visual contrast} strategy \cite{Jessica:2011} to highlight the busy places. Specifically, we assign blue and green, which are closed to the background color of the satellite map, to low values, in order to decrease their visual importance. On contrast, high values are encoded with yellow and red with high visual contrast.

In addition, we have mainly considered two alternatives for the map, as shown in the Figure 2. Similar visualization is predominantly based on the dark scheme, because it often brings about an appealing Sci-Fi feeling \cite{Dylan:2016}. However, it lacks in sense of realistic especially for external visitors. Regarding this, we adopted a 3D model map similar with the Wi-Crowd system \cite{Binthaisong:2017:WiCrowd}. Nevertheless, we still received negative feedback on its lack of realism and thus engagement. Therefore, our final prototype is based on the satellite map, which could help audiences locate the position easily. 

\begin{figure}
	\centering
	\includegraphics[width=1\columnwidth]{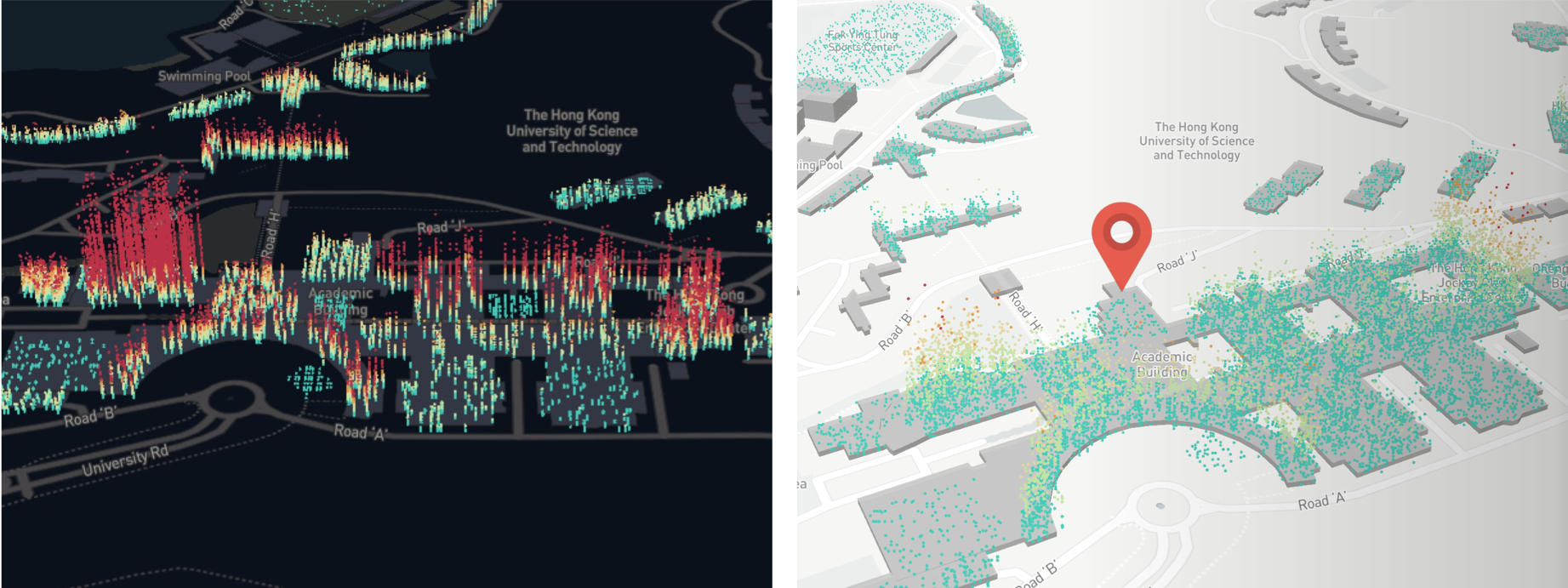}
	\caption{Two design alternatives for the map. Left: The dark map; Right: The 3D model map.}
\end{figure}

\subsubsection{Pop-up Window}
We designed a pop-up window (Figure 1.$A_1$) to display useful statistical information of important campus facilities in a loop, which a pin icon showing its location on the map. From top to bottom, it displays the place name with an icon showing its functionality, the current number of connection, the calculated level of crowdedness, a line chart showing the data of last 24 hours with highlights on peak timestamps.

We apply a \textbf{visual redundancy} method \cite{Jessica:2011} for showing the peak hours. Specifically, the line chart and peak hour table represent the same information but in different forms. We adopt both textual and visual forms to support different user preferences (\textbf{R2}).

\subsection{Chart View}
The Chart view (Figure 1.B) presents the statistical information including the crowd amount ranking of functional zones, the historical and predicted number, the movement of crowd among functional zones, and the buildings with top incoming and outgoing crowd.

We adopt simple, daily-used charts (i.e. bar and line charts) for the crowd ranking and total count (\textbf{R3}). During the design progress, we mainly compared two design alternatives (Figure 3). First, we found that the usage of gradient color as a \textbf{junk chart} component received more positive feedback, compared with uniform color scheme (Figure 3. Left). Second, while some audiences reported that the infographics with customized marks was interesting, this design was not preferred because it could be unfamiliar, non-intuitive and therefore take time to comprehend.

In order to visualize the campus movement, we used the standard approach in the visualization community - the Chord diagram. However, we received many complainants from the audiences during the prototyping stage, since they found it difficult to understand. We utilized the \textbf{anchoring} and \textbf{redundancy} techniques \cite{Jessica:2011} in response. Specifically, we calculated and set the top ten movement path as anchoring points, at which we highlighted the corresponding edge. Meanwhile, we provided textual information of that edge as redundancy. We got positive feedback that these improvements made it easier to understand.

We additionally designed a ladder to display the building with most incoming and outgoing crowd (\textbf{R4}). This idea was inspired at the initial interview where four audiences mentioned that they were interested in such ranking data. We also showed it to other audiences during the prototyping stage and received positive feedback.

\begin{figure}
	\centering
	\includegraphics[width=1\columnwidth]{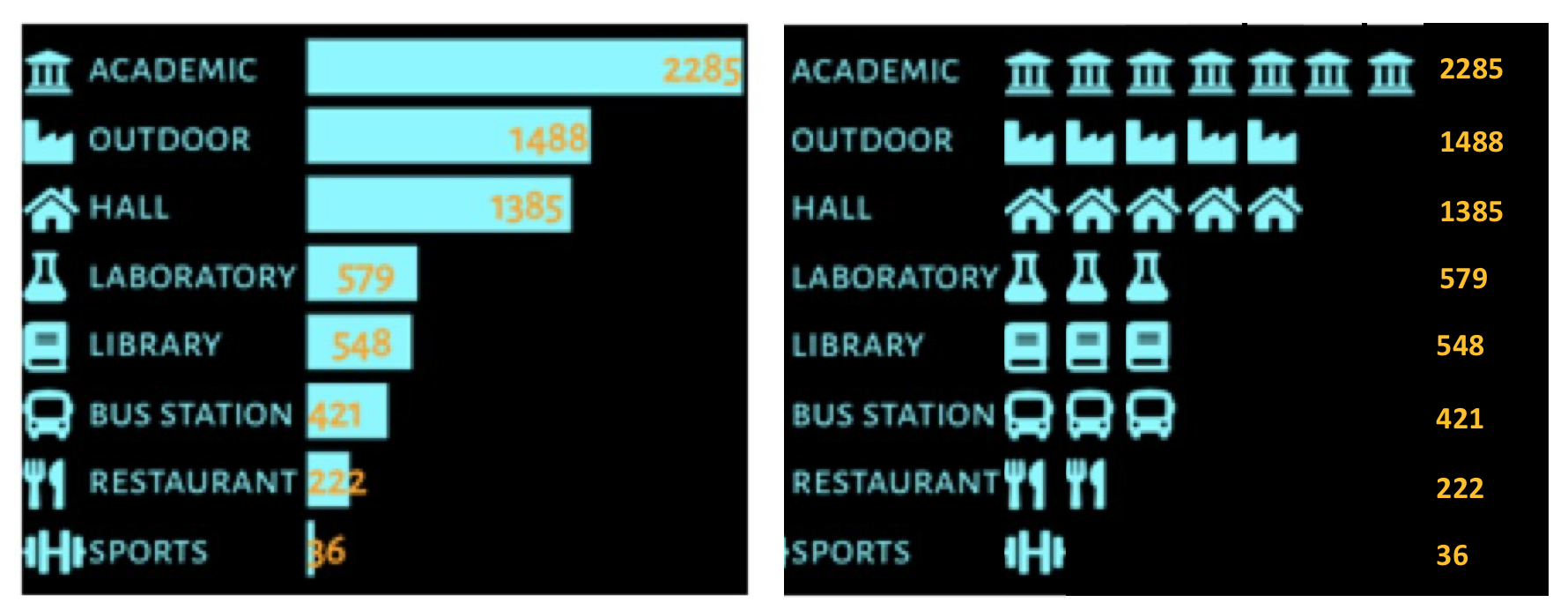}
	\caption{Two design alternatives for the bar chart. Left: Uniform color; Right: Infographics with customized marks.}
\end{figure}

\section{Evaluation}
Our third contribution is a user study to evaluate the benefits and attractiveness of the visualization design. Each user study session includes a two-minute demonstration, a questionnaire, and a five-minute interview. There are totally 33 participants (18 males, 15 females) between the age of 19 and 39 (mean 24.94, std 4.937). Those participants are excluded from the design progress. 

Figure 4 shows the description and results of questionnaires. As a whole, the visualization system receives an average score of 6.14, suggesting that the participants agree that it is useful, clear, interesting and engaging. However, participates reported less sanctification with the clearness. During the interview, two participants reported that it was still difficult to understand the Chord diagram. This indicates the needs for more effective methods for such communicative visualization. 

\begin{figure}
	\centering
	\includegraphics[width=1\columnwidth]{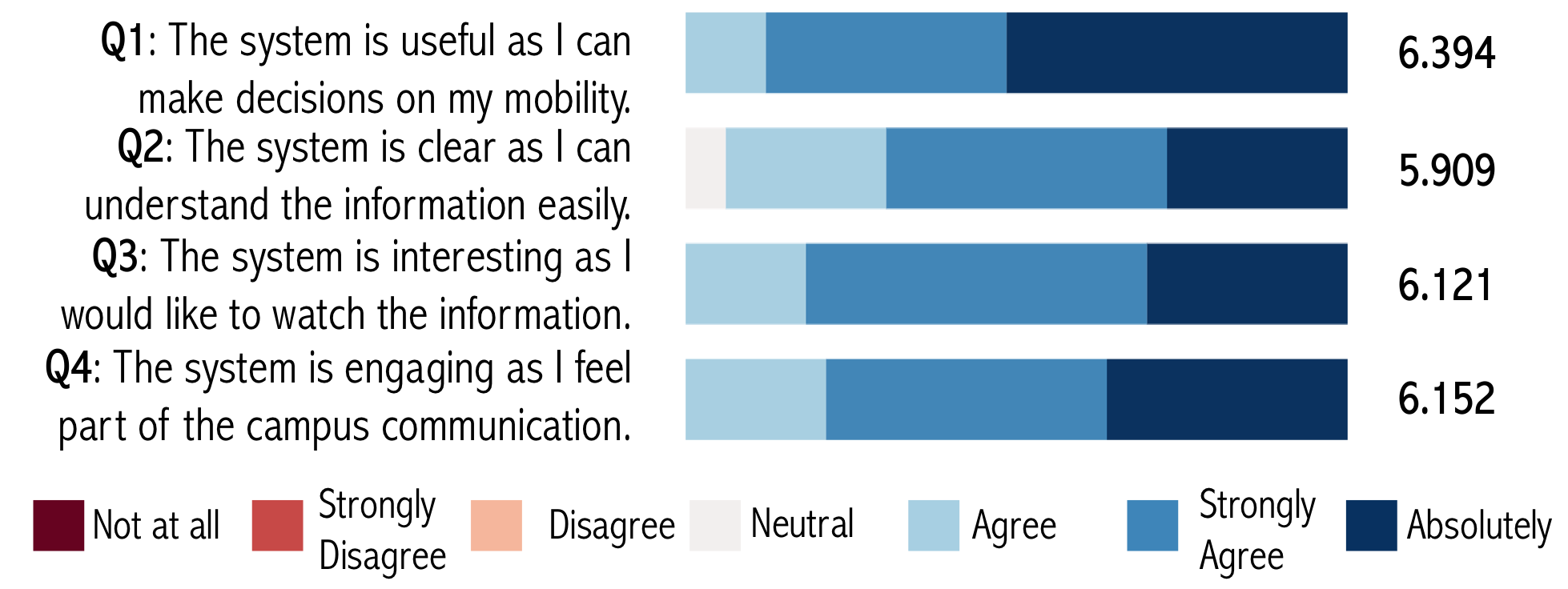}
	\caption{Description and results of questionnaires. The rightmost column denotes the average score using 7-Likert scale.}
\end{figure}

\section{Discussion}
Our work has two major limitations. First, our design decisions were mainly driven by anecdotal evidence gathered during the design progress, which might be biased. We plan to conduct statistical experiments to verify those assumptions. Second, our evaluation methodology was based on questionnaires and interviews. We intend to measure the success of the communication for more systematic evaluation in the future.

\section{Conclusion}
We report on our design progress of the Pulse system, a communicative visualization for crowd analytic with Wi-Fi access data on campus. During an iterative design progress, we determined the user preferences on different visual designs, and distilled four rationales that guide the design of such visualization system. The final prototype combines different visualization techniques such as visual contrast, visual redundancy, and junk chart to enhance the effectiveness of data communication. Feedback gathered from a user study indicates the benefits and attractiveness of our approach. We believe that our work could contribute to a growing scholarship on this application area.

\acknowledgments{We would like to thank Diana Liu, Justin Kwok, Tiffany TANG, and 	Serena YUNG for their valuable support and insight.}

\bibliographystyle{abbrv-doi}

\bibliography{poster}
\end{document}